\documentclass{JHEP3}

\usepackage{epsfig}
\usepackage{amssymb}
\usepackage{amscd}
\usepackage{amstext}
\usepackage{amsmath}
\usepackage{graphics}
\usepackage{amsfonts}
\usepackage{cases}
\usepackage{multirow}

\newcommand{\be}{\begin{equation*}}
\newcommand{\ee}{\end{equation*}}
\newcommand{\bea}{\begin{eqnarray}}
\newcommand{\eea}{\end{eqnarray}}
\newcommand{\bean}{\begin{eqnarray*}}
\newcommand{\eean}{\end{eqnarray*}}

\newcommand{\cf}{conifold }

\newcommand{\ig}{\includegraphics}

\newcommand{\tr}{\mathrm{Tr}}

\newcommand{\ag}{\alpha}
\newcommand{\bg}{\beta}
\newcommand{\cg}{\gamma}

\newcommand{\eg}{\varepsilon}

\newcommand{\agp}{\dot{\ag}}
\newcommand{\bgp}{\dot{\bg}}

\def\mD{\mathcal{D}}
\def\mF{\mathcal{F}}

\def\mJ{\mathcal{J}}
\def\mN{\mathcal{N}}
\def\mM{\mathcal{M}}
\def\mT{\mathcal{T}}
\def\mS{\mathcal{S}}

\def\mW{\mathcal{W}}

\def\IZ{\mathbb{Z}}

\def\IR{\mathbb{R}}

\def\IC{\mathbb{C}}
\def\IP{\mathbb{P}}

\def\mc{$\mM_C \phantom{i}$}
\def\mv{$\mM_V \phantom{i}$}

\def\beq{\begin{equation}}
\def\eeq{\end{equation}}

\preprint{MIT-CTP-3597\\  {\tt hep-th/0502043}}

\title{Conformal Manifolds for the Conifold\\ and other Toric Field Theories}

\author{Sergio Benvenuti$^1$, $\,$ Amihay Hanany$^2$\\

\vspace{0.3 cm}

1. \parbox[t]{6in}{Scuola Normale Superiore, Pisa,\\
                   and INFN, Sezione di Pisa, Italy.}

\vspace{0.5 cm}

2. \parbox[t]{6in}{Center for Theoretical Physics,\\
                   Massachusetts Institute of Technology,\\
                   Cambridge, MA 02139, USA.}\\

\email{sergio.benvenuti@sns.it, hanany@mit.edu}
}

\abstract{
In the space of couplings of the 4D $\mN = 1$ gauge theory associated to D3 branes probing Calabi-Yau singularities, there is  a manifold over which superconformal invariance is preserved. The AdS/CFT correspondence is valid precisely for this ``conformal manifold". We identify the conformal manifold for all the $Y^{p, q}$ toric singularities, paying special attention to the case of the conifold, $Y^{1, 0}$. For a general $Y^{p, q}$ the conformal manifold is three dimensional, while for the conifold it is five dimensional. There is always an exactly marginal deformation, analogous to the $\bg$--deformation of $\mN = 4$ SYM, which involves fluxes in the dual gravity description. This $\bg$--deformation exists for any toric Calabi-Yau singularity.}

\begin{document}

\section{Introduction and summary}
\label{section_introd}


Given a conformal field theory, a natural question that can be
asked is whether there is a continuos deformation that preserves
conformal invariance, called exactly marginal deformation. In
other words the question is if the conformal point is isolated in
the space of the couplings of the theory or if it is part of a
manifold of fixed points, which can be called the moduli space of
conformal field theories, $\mM_C$. In general one does not expect
to find such a manifold. The reason is that a fixed point is
defined by the vanishing of all the beta functions, and there is
one relation for each coupling, so in generic situations the fixed
points should be isolated. In supersymmetric theories the
situation can be different \cite{leighstrassler}; as is well known
non renormalization theorems imply that the beta functions for the
couplings can be expressed in terms of the anomalous dimensions of
the fundamental chiral fields. This is true both for the gauge
couplings and the superpotential couplings: the beta function is
proportional to a linear combination of the anomalous dimensions.
In this case, if there are more marginal couplings than anomalous
dimensions, one expects a conformal manifold \mc to exist.

The previous discussion for \mc is analogous to the problem of
finding a moduli space of gauge inequivalent vacuum expectations
values \cite{Kol:2002zt}: for non supersymmetric theories one
expects all these flat directions to be lifted by renormalization
effects, but in the supersymmetric case the non renormalization
theorem implies that for a large class of theories there is a non
trivial moduli space of vacua \mv. That these two objects share
somewhat similar properties can also be inferred by the AdS/CFT
correspondence \cite{Maldacena:1997re}: deforming a 4D CFT which has an AdS dual by an
exactly marginal operator corresponds to giving a vev to a
massless scalar field in the dual $AdS_5$ gravity description, so
\mc of the four dimensional theory should be the \mv of the 5D
theory \cite{Kol:2002zt}.

As is well known supersymmetric moduli spaces of vacua are computed
by
 \begin{itemize}
\item imposing $\mF$-term constraints on the complex
scalars of the theory, which are given by the chiral fundamental
fields,
\item imposing the $\mD$-term constraints,
\item modding out by the gauge transformations.
 \end{itemize}
 This procedure is equivalent to
imposing only the $\mF$-term constraints and to mod out by the
complexified gauge group, or to impose $\mF$-term constraints on
the basic gauge invariant operators. A similar procedure 
is valid also for the determination of the manifold $\mM_C$ in the following steps: 
\begin{itemize}
\item restrict the attention to chiral scalar
operators with R--charge $2$ (the ones that can enter the
superpotential without breaking the R--symmetry). Out of these only the ones entering the chiral
ring, that satisfy the $\mF$-term constraints, have protected scaling
dimension (in this case $3$) and are marginal operators,
 \item impose the real $\beta$-function relations on the
R--charges of the fundamental chiral fields,
 \item mod out by the
global symmetries broken by the deformations: a
field redefinition belonging to the broken symmetry group leaves
the K\"ahler potential of the theory invariant, and can be used to
put the chiral operators in a similar but different form.
\end{itemize}
Also for $\mM_V$, instead of imposing real $\beta$-function relations and modding out by the real global symmetry group, it should thus be possible to mod out by the complexified global symmetry group \cite{Kol:2002zt, aharony2002}.

We note that the scaling dimensions of all the chiral operators are constant on $\mM_V$. This is due to the constancy of the R--charges of the fundamental chiral fields, which is itself a simple consequence of the general field theoretic technique of $a$-maximization \cite{intriligator03} (the R--charges are solutions of quadratic equations with rational coefficients and cannot depend on continuos moduli). The number of chiral operators instead can change on $\mM_V$: when the superpotential changes, the $\mF$-term relations are different.

\vspace{0.1 cm}

Let us consider the well studied case of $\mN = 4$ SYM. As is well
known, there is a line of fixed points preserving $\mN = 4$ SUSY
and containing the free theory. Leigh and Strassler \cite{leighstrassler} showed the existence of two $\mN = 1$ marginal deformations, that break the $SU(3)$ global symmetry (this is the evident flavor symmetry of $\mN = 4$ in $\mN = 1$ language). One deformation preserves the Cartan generators of the $SU(3)$, and is called \mbox{$\bg$--deformation}\footnote{This deformation is called by various names in the literature. In the last period the term \mbox{$\bg$--deformation} \cite{Dorey:2003pp, Dorey:2004iq, Hollowood:2004ek, Benini:2004nn, Dorey:2004xm} seems to be the most popular.}, the other one breaks all the continuos flavor symmetries. In \cite{aharony2002} the corresponding deformations of the $AdS_5 \times S^5$ dual Type IIB supergravity have been computed, pertubatively in the deformations. Moreover, it has been shown that the space of marginal $\mN = 4$ SUSY breaking deformations corresponds to the two complex-dimensional quotient of the marginal chiral operators (which transform in the $10_\IC$ of $SU(3)$) by the complexification of $SU(3)$. 

\vspace{0.3 cm}

In this paper we want to study the exactly marginal deformations of superconformal field theories obtained by placing a stack of D3 branes at toric singularities. We will first analyze in detail the case of the conifold. The gauge theory was constructed in \cite{kw} and it gives one the most interesting extensions of the Maldacena conjecture, since it is one of the few cases for which the metric of the transverse space is explicitly known, for a review see \cite{conifoldrev}. It is also possible to deform this theory by adding fractional branes \cite{KNekrasov, KT, klebanovstrassler}. This deformation breaks the conformal invariance and gives the possibility of studying properties such as confinement through the interesting mechanism of Duality Cascade.

As already said, marginal deformations are reflected in the $AdS_5$ supergravity to vevs of massless scalars. The $AdS_5$ supergravity is a Kaluza-Klein reduction of ten dimensional Type IIB supergravity on \mbox{$AdS_5 \times X_5$}, where $X_5$ is a five dimensional compact manifold that, in the absence of fluxes, is Einstein (in order to preserve supersymmetry it has to be Sasaki-Einstein). In the ten dimensional Type IIB supergravity a marginal deformation corresponds to a continuos modification of the expectation value of all the bosonic tensor fields, involving only the compact manifold $X_5$.

\vspace{0.1 cm}

In section \ref{section_margin} we apply the procedure described above to the conifold field theory. We find that \mc has complex dimension $5$. One deformation is associated to the difference between the field strengths of the two gauge groups, while the other four are related to superpotential terms. One term is the Klebanov-Witten superpotential discussed in \cite{kw}, it preserves the largest possible flavor symmetry, $SU(2) \times SU(2)$. There is also one term which breaks only half of the flavor symmetry; for this term the corresponding supergravity background is known \cite{orbifoldsugra}, and was discussed also in \cite{corrado04}; it is called the Pilch-Warner superpotential. The other two terms, breaking the flavor symmetry further, are also written explicitly.

\vspace{0.1 cm}

In section \ref{secYPQ} we extend the analysis to a family of superconformal field theories that generalize the case of the conifold, called $Y^{p, q}$, recently constructed in \cite{Benvenuti:2004dy}. This set is interesting because the explicit metric on the compact Sasaki--Einstein manifold is known, thanks to the important results of \cite{GMSW, Martelli:2004wu}. The $Y^{p, q}$ field theory is a superconformal quiver with precisely $2 p$ gauge groups. The conifold, with $2$ gauge groups, corresponds to $Y^{1, 0}$. An important feature of these geometries is that they are toric \cite{Martelli:2004wu} \footnote{A nice exposition of toric geometry, and in particular toric Calabi--Yau singularities, can be found in \cite{Martelli:2004wu}.}, and the so called \emph{quiver toric conditions} \cite{Feng:2000mi, Feng:2002zw} become useful in the study of the theories. Also for this set of models fractional branes can be added. The authors of \cite{Herzog:2004tr} were able to explicitly find the exact asymptotic supergravity background with the corresponding \mbox{$3$--form} fluxes turned on. In \cite{Herzog:2004tr} various supergravity computations have been matched with field theoretical properties of the cascading theories. A difference with respect to the conifold is that for any given geometry there is a corresponding infinite set of conformal theories, all related to each other by Seiberg Dualities \cite{Seiberg:1994pq}. Some of these phases satisfy the toric condition, i.e. all the gauge groups have the same rank. In \cite{Benvenuti:2004wx} the set of the toric phases has been described. 

For a general $Y^{p, q}$ the conformal manifold is three complex dimensional. Precisely one exactly marginal deformation involves fluxes in the dual gravity description. It is the analog of the $\bg$--deformation of $\mN = 4$ SYM. This $\bg$--deformation exists for any toric Calabi--Yau singularity. For any toric quiver, each bifundamental field appears in the superpotential precisely twice, once with a positive sign and once with a negative sign. The \mbox{$\bg$--deformation} is generated by this toric superpotential, where all the `minus' signs are switched to `plus' signs. An immediate consequence is that the Cartan generators of the global flavor symmetries are symmetries also of the $\bg$--deformed theory. For generic toric quivers the undeformed symmetry has a subgroup $U(1)^2_F \times U(1)_R$. This subgroup is preserved by the $\bg$--deformation. As a consequence of this and of the quotient procedure described above, the $\bg$--deformation is always exactly marginal.\footnote{It is clear that the operator that drives the $\bg$--deformation lies in the chiral ring, since $\mF$--term relations of the underformed theory always contain both `plus' and `minus' signs.}

\vspace{0.1 cm}

In section \ref{section_conclu} we comment on the interesting possibility of finding the supergravity dual of the $\bg$--deformation.

\vspace{0.4 cm}

A change in the superpotential of the theory implies a
modification of the moduli space of vacua $\mM_V$. In section
\ref{section_moduli}, limiting the analysis to the case of the conifold, we show that \mv depends on the
superpotential couplings only modulo the action of the
complexified flavor symmetry group, which becomes $SO( 4, \IC )$.
The stable orbits of this action correspond, as expected, to the
exactly marginal deformations discussed in section
\ref{section_margin}. If only the Klebanov-Witten superpotential
is present \mv is the full $3$-complex dimensional conifold; the
low energy Higgsed theory, corresponding to taking all the
coincident D3 branes out of the singularity is the $\mN = 4$
theory, as expected from the fact the no fluxes are turned on \cite{kw, Klebanov:1999tb}.
Adding the Pilch-Warner term lifts one marginal direction and the
moduli space of vacua \mv becomes the singularity $\IC^2/\IZ_2$; in this case if we move the branes outside the singularity as in \cite{Klebanov:1999tb} we find the $\mN = 1^*$ theory (this is the infrared superconformal theory obtained by adding one supersymmetric mass term to the $\mN =4$ theory, also called the Leigh--Strassler fixed point).
Also this result is expected, since now the background has fluxes
turned on. If all the possible terms are turned on we do not find
any marginal direction (\mv reduces to a point), but there are
particular combinations of the exactly marginal superpotentials
that leave exactly one marginal direction; giving a vev to the
fundamental fields and flowing to the infrared one obtains the
$\mN = 2^*$ theory (corresponding to $\mN = 4$ plus $\mN = 2$ mass
terms). All of these results are consistent with the fact the
superpotential studied are exactly marginal and that on the string
side they correspond to continous deformations involving fluxes.

\vspace{0.4 cm}

In the previous discussion we did not consider the unstable orbits
of the $SO( 4, \IC)$ action. In appendix \ref{nondiago} these are classified. By the analogy with
$\mM_V$, we do not expect these superpotentials to be exactly
marginal. This expectation is also confirmed by the fact that
Leigh--Strassler arguments do not work for these superpotentials.
After having computed $\mM_V$ for these deformations, one finds in
all the possible cases the following situation: giving a vev to the fields and flowing to the IR, the resulting IR Higgsed theory has a $\mM_V^{IR}$ whose dimension is larger than the dimension of the original $\mM_V$.
This is a contradiction, and we interpret this fact as an evidence
that unstable superpotentials are marginally irrelevant operators. As a consequence, the conformal manifold is stable under marginal perturbations. This feature is espected to be of some relevance in the study of initial conditions dependence in Duality Cascades.

\newpage
\section{The conformal manifold of the conifold field theory}\label{section_margin}
In this section we describe all the possible supersymmetric
marginal deformations of the interacting superconformal field
theory associated to the conifold geometry. Then we isolate the
space of exactly marginal deformations, showing also that it can
be thought of as the space of all the marginal deformations modded
out by the complexified global symmetry group broken by the
couplings.


\subsection{Review of the field theory and marginal deformations}
The conifold field theory has a product gauge group $U(N) \times
U(N)$. The matter chiral superfields live in the bifundamental
representations of this gauge group: there are two fields ($A_1$
and $A_2$) transforming in the $(N, \overline{N} )$ and two fields
($B_{\dot1}$ and $B_{\dot{2}}$) transforming in the $( \overline{N}, N )$. This
matter content can be encoded in the following quiver diagram.

\begin{figure}[h]
  \epsfxsize = 7.2 cm
  \centerline{\epsfbox{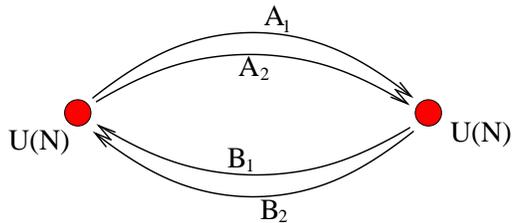}}
  \caption{The quiver diagram of the conifold field theory.}
  \label{quiverdiagram}
\end{figure}
This quiver diagram is known \cite{kw} to arise from the low
energy excitations of a stack of $N$ D3 branes placed at the
singular point of the conifold geometry. The conifold is a
$3$-complex dimensional manifold defined by the following
quadratic equation in $\IC^4$: 
\beq \label{conifoldwi} w_1^2 + w_2^2 + w_3^2 + w_4^2 = 0\,. \eeq 
This space also represents the
moduli space of vacua \mv of the gauge theory, as it corresponds
to the configuration space of $N$ parallel D3 branes (more
precisely the moduli space is the symmetrized product of $N$
conifolds). There is an isolated singularity at $w_1 = w_2 = w_3 =
w_4 = 0$.

The \cf can be viewed as a real cone over the compact $5$-real
dimensional manifold called $T^{1, 1}$. As clearly explained in
\cite{kw}, $T^{1, 1}$ can be explicitly described as the manifold
$SU(2) \times SU(2) /U(1)$, has the topology of $S^2 \times
S^3$ and can be seen as a $U(1)$ fibration over the regular K\"ahler--Einstein manifold $\IC\IP^1 \times \IC\IP^1$.

In order to completely identify the $4$D superconformal gauge
theory it is necessary to specify the superpotential $\mathcal{W}$. In \cite{kw} this was argued to be
\beq \label{Wkw} \mathcal{W}_{KW} \propto \varepsilon^{\ag \bg}
\varepsilon^{\agp \bgp} \, \tr \left( A_{\ag} B_{\agp} A_{\bg}
B_{\bgp} \right)\,. \eeq 
This is the only possible superpotential
consistent with superconformal invariance that preserves a global
$SU(2) \times SU(2)$ flavor symmetry. One $SU(2)$-factor acts on
the fields $A_1$ and $A_2$, which transform as a
doublet, the other $SU(2)$-factor acts on the doublet $B_{\dot1}$ and
$B_{\dot{2}}$. These global symmetries are associated to the isometries
of $\IC\IP^1 \times \IC\IP^1$. Since $T^{1, 1}$ is a
$U(1)$-fibration over $\IC\IP^1 \times \IC\IP^1$ there is another
symmetry corresponding to rotations of the fiber: this is the
$U(1)_R$-symmetry which at the superconformal fixed point is the
non-anomalous partner of the stress-tensor; it acts on the $4$
bifundamental fields with charge $1/2$: in this way the
superpotential (\ref{Wkw}) has total R--charge $2$ and is a
marginal operator. Since the R--charge of a free chiral
superfield is $2/3$, the fact that the R--charges are $1/2$
implies that the superconformal field theory is a strongly
interacting gauge theory, ``far'' from the free theory where the
R--charges of the bifundamental fields are $2/3$. The last
continuos global symmetry present in the gauge theory is a
baryonic $U(1)$ symmetry, acting with charge $+ 1$ on the two fields
$A_\ag$ and charge $- 1$ on the fields $B_{\agp}$.\footnote{This
baryonic $U(1)$ can be seen as part of the gauge group $U(N)
\times U(N)$: out of the two $U(1)$ factors, a diagonal
combination is fully decoupled (no bifundamental field is charged
under it), and the other diagonal combination decouples in the IR
and can be identified with this baryonic $U(1)$ global symmetry.}
This baryonic symmetry is not associated to isometries of $T^{1,
1}$, but to the presence of the $S^3$ cycle in the topology of
$T^{1,1}$. 

Our purpose is to study the strongly coupled field theory
described by the quiver diagram of Figure \ref{quiverdiagram} with a generic
superpotential that preserves conformal invariance. This corresponds
precisely to the set of theories for which an AdS string dual
exists. Of course, for a generic point on the conformal manifold, all the bosonic Type IIB fields are turned on -- not just the metric and five form, and the dual supergravity background is not known. We should also remark that this IR-conformal manifold can arise from different UV descriptions. Maybe the more interesting UV description is \cite{kw} the quiver associated to the orbifold $\IC \times \IC^2/\IZ_2$, since the full RG flow (the Klebanov--Witten flow), driven by twisted sector mass terms, has a dual gravity description \cite{kw, Halmagyi:2004jy}.

The single-trace marginal superpotentials are precisely the
polynomials quartic in the bifundamental fields \footnote{We are
assuming $N \geq 5$: for $N =2$ or $N = 3$ some terms becomes degenerate, while for $N = 4$ it is
possible to consider also ``baryonic'' superpotentials, of the
form $\eg \, A \, A \, A \, A \, \eg$.}, of the following form:
\beq \label{genericW} \nonumber \mathcal{W}_{gen} = \lambda^{\ag \bg , \;
\agp \bgp} \, \tr \left( A_{\ag} B_{\agp} A_{\bg} B_{\bgp}
\right)\;\,. \eeq 
Some of these deformations will be exactly
marginal, the other ones will be argued in section \ref{section_moduli} and in appendix \ref{nondiago}
to be marginally irrelevant. The $A_i$ transform in the
representation $(\frac{1}{2}, 0)$  of the group $SU(2) \times
SU(2)$, while the $B_{j}$ in the $(0, \frac{1}{2})$, so a generic
quartic term will transform as \beq \left( \frac{1}{2}, 0 \right)
\otimes \left(0 , \frac{1}{2}\right) \otimes \left( \frac{1}{2}, 0
\right) \otimes \left(0 , \frac{1}{2}\right) = (0, 0) \oplus (1,
0) \oplus (0, 1) \oplus (1, 1)\;. \eeq 

The trace in front of the monomials annihilates the terms $(1, 0) \oplus (0, 1)$, so the most general single trace superpotential is
 \beq \label{genericW2} \mathcal{W}_{gen} = \lambda_{KW}
\varepsilon^{\ag \bg} \varepsilon^{\agp \bgp} \, \tr \left( A_{\ag}
B_{\agp} A_{\bg} B_{\bgp} \right) + \lambda^{(\ag \bg) , \; (\agp
\bgp)} \, \tr \left( A_{\ag} B_{\agp} A_{\bg} B_{\bgp} \right)\;,
\eeq
 where $\lambda^{(\ag \bg) , \; (\agp \bgp)}$ is symmetric in
the indices $\ag, \bg$ and $\agp, \bgp$.

In total there are $10$ marginal superpotential terms: one $SU(2)
\times SU(2)$ preserving term and nine $SU(2) \times SU(2)$
violating terms. All the couplings associated to these
deformations are complex. It is important to note that all the operators in the $(1, 1)$--spin are chiral operators, i.e. satisfy the $\mF$--term constraints arising from the superpotential in (\ref{Wkw}). They are part of the series of chiral mesons with spin $(k/2, k/2)$ and R--charge $k$ discussed in \cite{kw}. 

\subsection{The five dimensional conformal manifold}
The remaining part of this section is devoted to explain at some
length the structure of the RG flows for the gauge theory
previously described.

Let us start from the free theory associated to the quiver diagram
of Figure \ref{quiverdiagram}: the two gauge couplings and all the superpotential
couplings $\lambda^{\ag \bg , \; \agp \bgp}$ are zero. Clearly the
superpotential couplings are strictly irrelevant, since the
superpotential is quartic.
The two gauge couplings are instead relevant, since
each gauge group can be seen as a $SU(N_C)$ SQCD with $N_F = 2
N_C$ flavors. The exact $\mathcal{N} =1$ beta function for the
gauge couplings, in terms of the anomaluos dimensions of the
fundamental chiral fields $\cg(g)$, is:
 \beq\label{NSVZa}  \beta_{\frac{1}{g^2}} =  \frac
    {3 N_C - \sum_{M} \mu[\mathcal{R}_M] (1 - 2 \cg_{M}(g))}{8 \pi^2 - g^2 N}\;,
\eeq
 where $\mu[\mathcal{R}]$ is the Dinkin index of the
representation $\mathcal{R}$, defined by $Tr_{\mathcal{R}}T^a T^b
= \mu[\mathcal{R}] \delta^{a\,b}$. In the case of the fundamental
representation $\mu = \frac{1}{2}$. For our purpose it is enough
to take in consideration the numerator, since the rest is a
function of the couplings that does not change sign along the
Renormalization Group flow. So (\ref{NSVZa}) becomes
\beq\label{NSVZb} \bg_g \propto - \left( N + \frac{N}{2} \sum_{M}
(r_M - 1) \right) \propto - \left( r_{A_1} + r_{A_2} + r_{B_{\dot1}} +
r_{B_{\dot{2}}} - 2 \right)\;, \eeq where we changed variables, from the
anomalous dimensions $\cg$ to the R--charges, using the relation
between the total scaling dimension $D$ of a chiral operator and
its R--charge $D = 1 + \cg  = 3 / 2 \, r$. The R--charges are
functions of all the couplings of the theory, in this case the two
gauge couplings $g_1$, $g_2$
and the superpotential couplings $\lambda^{\ag \bg , \; \agp \bgp}$.\\
The beta function for the superpotential couplings are instead
proportional to the total R--charge of the monomials minus $2$:
\beq\label{betaW} \beta_\lambda \propto  \sum_M r_{M} - 2 \;, \eeq
where the sum is over the fields entering in the monomial. We note that (\ref{betaW})
can be different from (\ref{NSVZb}), for example, in the case $\mathcal{W} = \lambda \tr
\left( A_{1} B_{\dot{1}} A_{1} B_{\dot{1}} \right)$, (\ref{betaW}) is $\beta_\lambda
\propto  2 r_{A_1} + 2 r_{B_{\dot1}} - 2$. Near the free fixed point, where all R--charges are close to
$2/3$, (\ref{NSVZb}) is indeed negative and (\ref{betaW}) is
positive.

Now we can follow the RG flow driven by one of the two gauge
couplings, say $g_1$, keeping the other couplings zero. At the end
there is a IR fixed point where $\bg_{g_1}$ vanishes. This fact
imposes one relation on the four R--charges of the bifundamental
fields: their sum has to be $2$. It is clear that the result is
that all the R--charge are $1/2$, since the four bifundamental
fields enter symmetrically.

Now we can add other deformations to the theory. Considering the
other gauge coupling ($g_2$), the constraint given by $g_2$ is
always (\ref{NSVZb}), the same as $g_1$. In other words the RG
flow of the theory without superpotential looks like:

\begin{figure}[h]
  \epsfxsize = 5.5 cm
  \centerline{\epsfbox{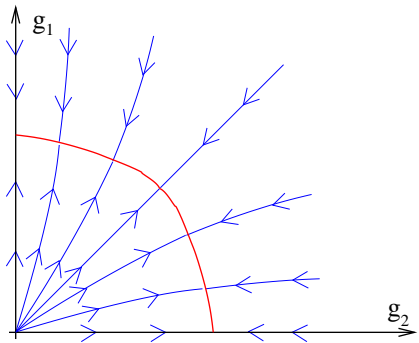}}
  \caption{The fixed line with vanishing superpotential.}
  \label{flowcon1}
\end{figure}

In the plane $g_1, g_2$ there is thus a line of superconformal
fixed points. This line is characterized by the condition \beq
r_{A_1} = r_{A_2} = r_{B_{\dot1}} = r_{B_{\dot{2}}} = \frac{1}{2} \;. \eeq It is
worth noting that these $4$ conditions effectively give only one
constraint on the values of the two gauge coouplings $g_1$ and
$g_2$. On this fixed line the chiral ring scalar operators with
R--charge $2$ are the $10$ operators of (\ref{genericW2}).

At this point we can study deformations given by a superpotential.
A quartic term is irrelevant at the free fixed point, since its
total R--charge is strictly greater than $2$, but at the
interacting fixed line found above the total R--charge becomes
exactly $2$, so a quartic term is a candidate for an exactly
marginal operator.

Adding the Klebanov-Witten term gives a $\bg$-function relation
precisely of the same form as (\ref{NSVZb}). Since the global
symmetry is preserved, the two $A_\ag$ R--charges are still equal.
The same is true for the two $B_{\dot\ag}$ R--charges. A $\IZ_2$ symmetry
than implies that there is always just one free R--charge. Since
there is precisely one independent relation, we see that there is
one more marginal direction, with respect to Figure \ref{flowcon1}. In other words the three
complex couplings $g_1$, $g_2$ and $\lambda_{KW}$ break just a
$U(1)$ factor of the naive $U(2) \times U(2)$ global symmetry, so they
define a two--complex dimensional conformal manifold. On this fixed
surface the R--charge $2$ chiral operators are only the ones
transforming in the $(1, 1)$. As explained in \cite{kw}, these two marginal directions have a direct interpretation in supergravity. Increasing $\lambda_{KW}$, $g_1$ and $g_2$ with $1/g^2_1 - 1/g^2_2$ fixed corresponds to changing the vev of the complex Type IIB dilaton. Keeping $\lambda_{KW}$ fixed and changing the $1/g^2_1 - 1/g^2_2$ corresponds to changing the vev of the complex Type IIB B field on the $S^2$.

Adding now a symmetry breaking term in the $(1,1)$ of $SU(2) \times SU(2)$ generates
new $\bg$-function relations. Instead of analysing these relations
we argue the form of the marginal deformations by modding out the
space $(1,1)_\IC$ by $SU(2, \IC) \times SU(2, \IC)$. The space
$(1,1)_\IC$ is the space of $4 \times 4$ symmetric traceless
matrices, where the group $SU(2, \IC) \times SU(2, \IC) = SO (4,
\IC)$ acts as rotations. The stable orbits of the action are
clearly the diagonal traceless matrices. (In section \ref{section_moduli} the
various changes of variables are written explicitly). The result is
that the three following operators are exactly marginal
deformations:
 \beq \label{genericmarginal}
 \left\{
\begin{array}{l}
 \tr \left( A_{1} B_{\dot{1}} A_{2} B_{\dot{2}} + A_{1} B_{\dot{2}} A_{2} B_{\dot{1}} \right) \\
 \tr \left( A_{1} B_{\dot{1}} A_{1} B_{\dot{1}} + A_{2} B_{\dot{2}} A_{2} B_{\dot{2}} \right) \\
 \tr \left( A_{1} B_{\dot{2}} A_{1} B_{\dot{2}} + A_{2} B_{\dot{1}} A_{2} B_{\dot{1}} \right)\;.
\end{array}
\right. \eeq
 For generic values of the associated couplings the global flavor symmetry is
completely broken (only the baryonic $U(1)$ survives). It is
actually possible to verify that (\ref{genericmarginal}) is
exactly marginal using Leigh--Strassler type \cite{leighstrassler} arguments:\\
(\ref{genericmarginal}), as $\mathcal{W}_{KW}$, is invariant under a discrete $\IZ_2 \times \IZ_2$ symmetry. One factor exchanges $A_1$ with $A_2$ \emph{and} $B_{\dot1}$ with $B_{\dot{2}}$. The other factor changes the sign of $A_2$ and $B_{\dot{2}}$. This discrete symmetry implies that the $2 \times 2$ matrix of the anomalous dimensions for the chiral fields $A_1$ and $A_2$ is proportional to the identity (the same is true for $B_{\dot1}$ and $B_{\dot{2}}$). The two doublets of bifundamental fields enter symmetrically, so there is one undetermined R--charge (as in the previous case without superpotentials), whose value is still $1/2$.

In \cite{Ceresole:1999zs} the full Kaluza-Klein reduction of Type IIB supergravity on $T^{1, 1}$ was found. From this result one can read off the supergravity excitations corresponding to the operators (\ref{genericmarginal}); these massless excitations are turned on at first order in the deformation. At higher orders, however, we expect all the supergravity fields to be turned on.

The full set of exactly marginal superpotentials, (\ref{genericmarginal}) plus the Klebanov-Witten term, can be written as 
\beq \label{genericmarginal2}
\begin{array}{r l}
\mW = & \; \lambda_{KW} \, \tr \left( A_{1} B_{\dot{1}}A_{2} B_{\dot{2}} - 
            A_{1} B_{\dot{2}} A_{2} B_{\dot{1}} \right) \nonumber +\\
 & \lambda_{PW} \, \tr \left( A_{1} B_{\dot{1}} A_{2} B_{\dot{2}} + \nonumber
           A_{1} B_{\dot{2}} A_{2} B_{\dot{1}} -  A_{1} B_{\dot{2}} A_{1} B_{\dot{2}}
         - A_{2} B_{\dot{1}} A_{2} B_{\dot{1}} \right) +\\
 & \;\;\; \lambda_\bg \;\, \tr \left( A_{1} B_{\dot{1}} A_{2} B_{\dot{2}}  +
                     A_{1} B_{\dot{2}} A_{2} B_{\dot{1}} \right) +\\
 & \;\;\; \lambda_2 \;\, \tr \left( A_{1} B_{\dot{1}} A_{1} B_{\dot{1}} \nonumber
                + A_{2} B_{\dot{2}} A_{2} B_{\dot{2}} \right)\;.
\end{array}
\eeq
The term proportional to $\lambda_{PW}$ preserve an $SU(2)$ global symmetry and is called the Pilch-Warner superpotential. In \cite{orbifoldsugra, corrado04} it has been argued that the conifold field theory with the Pilch-Warner superpotential is an orbifold $\IZ_2$ of the $\mN = 1^*$. This theory is a massive deformation of the $\mN = 4$ SYM and has been studied holographically in \cite{Freedman:1999gp, Pilch:2000ej, Pilch:2000fu, Pilch:2004yg}. 

The term proportional to $\lambda_\bg$ is the analog of the so called $\beta$-deformation of $\mN = 4$ SYM. A similar term can be written down for any toric quiver and is the deformation that generalizes to the set of superconformal field theories discussed in the next section.


\section{The conformal manifold of the $Y^{p, q}$ quivers}\label{secYPQ}
In this section we consider a class of superconformal field theories that generalizes the conifold. They arise as the low energy excitations of D3 branes placed at a Calabi-Yau singular three-fold which is a cone over the so called $Y^{p, q}$ Sasaki-Einstein manifold \cite{GMSW}. The interesting thing about this class of models is that the explicit metric on the Calabi-Yau cone is known, thanks to the important results of \cite{GMSW}. The dual gauge theories, that have to be superconformal quivers, have been recently constructed in \cite{Benvenuti:2004dy}, following the algebro-geometric description of the $Y^{p, q}$ singularities given in \cite{Martelli:2004wu}. The quivers have precisely $2 p$ gauge groups; the conifold, with $2$ gauge groups, corresponds to $Y^{1, 0}$. An important feature of these geometries is that they are toric \cite{Martelli:2004wu}, this immediately tells that all the gauge groups have the same rank, thanks to the so called \emph{quiver toric condition} \cite{Feng:2000mi, Feng:2002zw}. The models $Y^{p, p}$ correspond to the known orbifolds $\IC^3/\IZ_{2 p}$, with action $(1, 1, - 2)$.
It also turns out \cite{Martelli:2004wu} that $Y^{2,1}$ correspond to the del Pezzo$_1$ singularity, so also this gauge theory is known: the quiver and the exact toric superpotential were given in \cite{Feng:2000mi}; this case has been discussed in \cite{Feng:2002zw, intriligator03, Bertolini:2004xf}.

All the $Y^{p, q}$ Sasaki-Einstein five manifolds share the same topology of the conifold: $S^2 \times S^3$. A difference with respect to the conifold is that for any given geometry there is a corresponding infinite set of conformal theories, all related to each other by Seiberg Dualities \cite{Seiberg:1994pq}. Some of these phases satisfy the toric condition, i.e. all the gauge groups have the same rank. In \cite{Benvenuti:2004wx} the set of the toric phases has been described. Seiberg Dualities also play a role when conformal invariance is broken by fractional branes and a Duality Cascade starts, and in \cite{Herzog:2004tr} they were studied in the presence of fractional branes. In this context, the exact asymptotic supergravity background with the corresponding $3$--form fluxes turned on has been explicitly found \cite{Herzog:2004tr}.\footnote{These deformed geometries can also be extended to more general sets \cite{Pal:2005mr}.}

The purpose of this section is to apply the procedure discussed in detail in section \ref{section_margin} to a general $Y^{p, q}$ theory. The quivers constructed in \cite{Benvenuti:2004dy} will be considered. Any Seiberg dual description would give precisely the same results. This is a direct consequence of the original results of \cite{Seiberg:1994pq} and of the fact that Seiberg Duality for quivers can be thought of as Seiberg Duality on a particular $SU(N)$ gauge theory with some flavors. Since all of our results belong to the protected sector (BPS operators, global symmetries, etc.), they are automatically invariant under Seiberg Duality.

For a detailed explanation of the $Y^{p, q}$ field theories we refer to \cite{Benvenuti:2004dy}, for an in depth exposition of many important geometric properties we suggest \cite{Martelli:2004wu}.

In \cite{Benvenuti:2004dy} the quiver diagrams of the $Y^{p, q}$ superconformal field theories have been found. In Table \ref{examplefig} we reproduce some examples, with the hope of being self-explanatory. 

\begin{table}[h]\begin{center}
$$\begin{array}{ccc}
{\ig[height=4.6 cm]{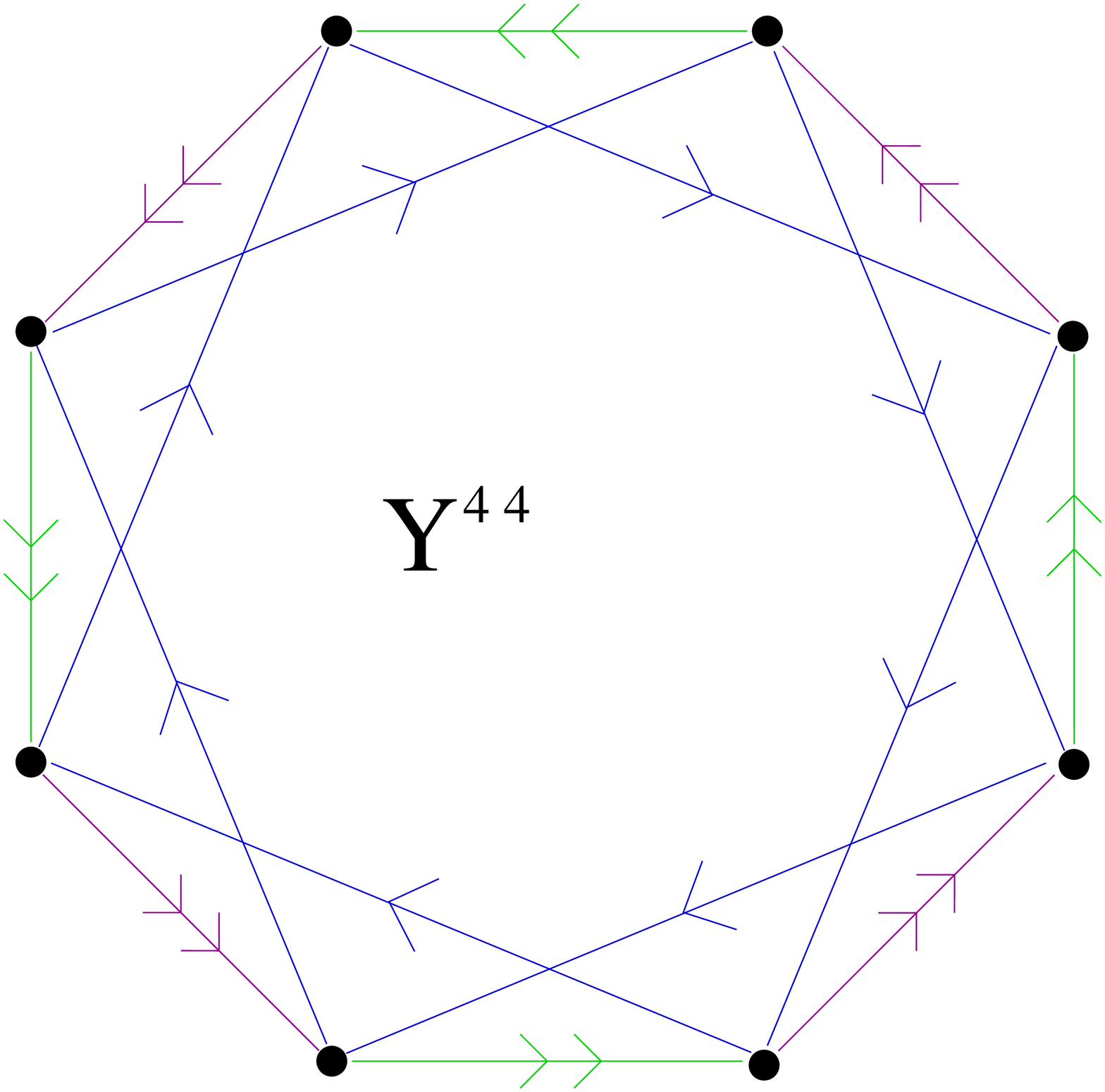}} \;\;\;\;\;  &
{\ig[height=4.6 cm]{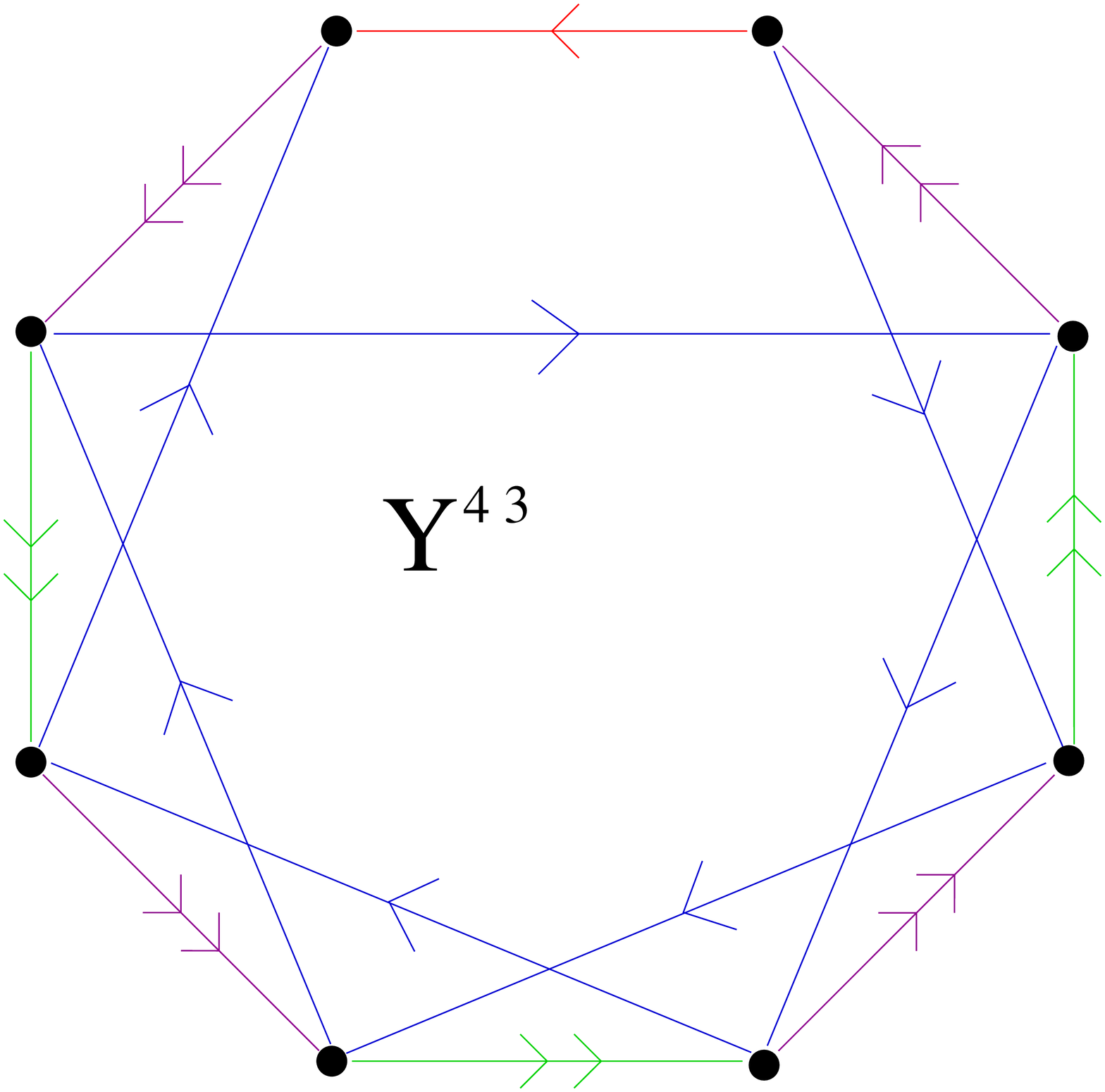}} \;\;\;\;\;  &
{\ig[height=4.6 cm]{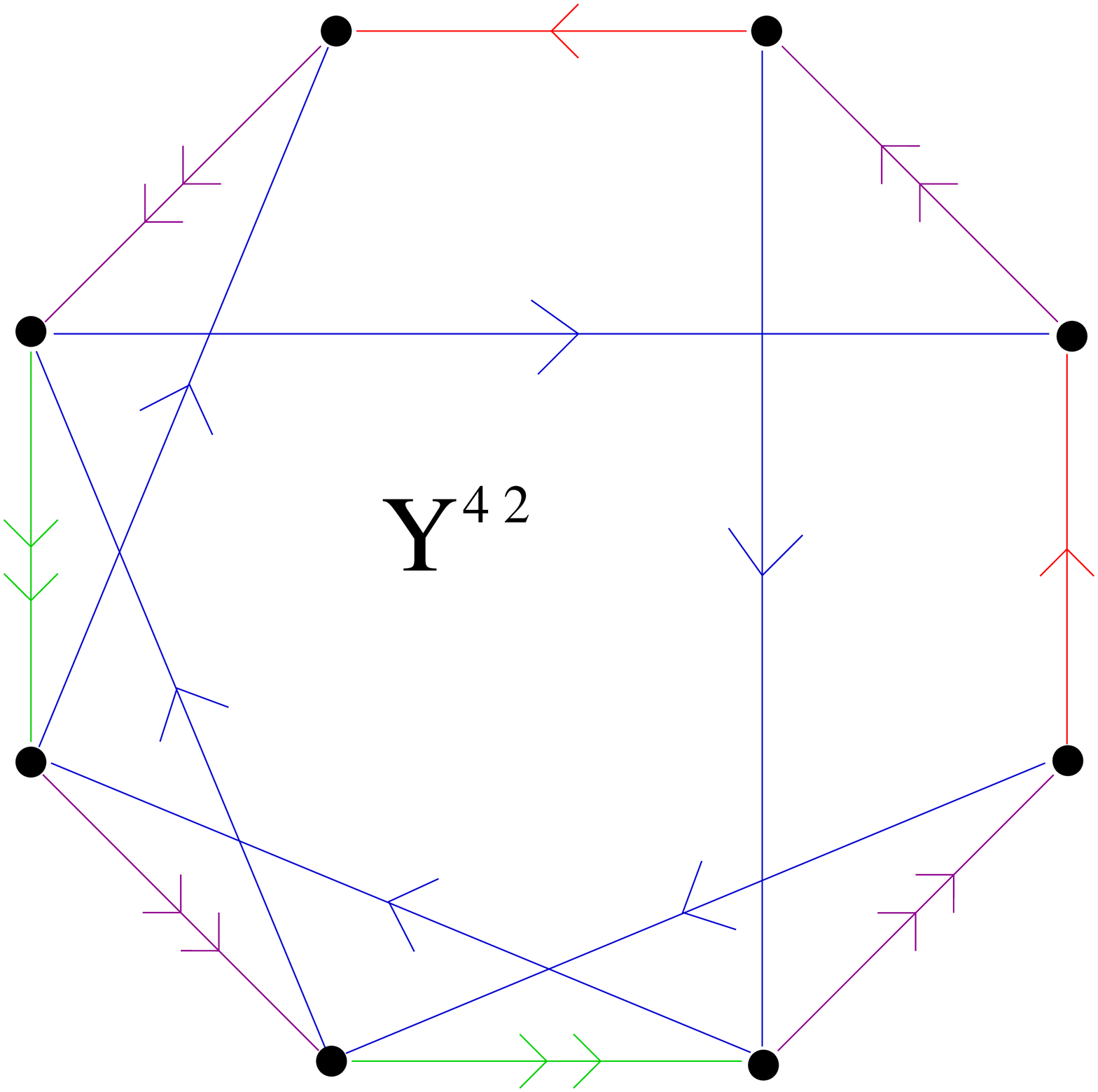}}  \end{array}$$
\caption{Example of the recursive construction of the $Y^{p, q}$ quivers. These figures are reproduced from \cite{Benvenuti:2004dy}.} \label{examplefig}
\end{center}\end{table}

The exactly marginal superpotential is \cite{Benvenuti:2004dy} (the sum over gauge indices is implicit)
 \beq\label{WYpq}
W = \sum_{i = 1}^q \epsilon_{\ag \bg}
( U_i^\ag V_{i}^\bg Y_{2i - 1} +  V_i^\ag U_{i+1}^\bg Y_{2i} )
    + \sum_{j = q + 1}^p \epsilon_{\ag \bg} Z_j U_{j+1}^\ag Y_{2j - 1} U_j^\bg\;.
\eeq
The fields $U$ and $V$ are in the spin--$1/2$ of the global $SU(2)$; $Y$ is the ``internal'' singlet and $Z$ is the ``external'' singlet.
It is possible to modify this superpotential. Keeping marginality and the $SU(2)$ symmetry, there are $2q$ parameters associated to the cubic terms of the form $\tr ( U V Y )$ and $p - q$ parameters associated to the quartic terms of the form $\tr ( U Z U Y )$. Moreover, there are the $2 p$ marginal gauge couplings. There are thus a total of $3 p + q$ marginal couplings.

In \cite{Benvenuti:2004dy} it was shown that the solution of the $3 p + q$ linear $\bg$--function contraints is a two dimensional space. One surprising aspect of the analysis of \cite{Benvenuti:2004dy} is that there is a strong degeneracy of R--charges: all the $Y$--fields have the same R--charges, all the $Z$--fields have the same R--charges etc. This is a direct consequence of the imposition of all the $3 p + q$ linear $\bg$--function relations.\footnote{It is interesting to note that, instead of analysing all the linear relations, one can assume the degeneracy among the R--charges and then keep track of the fact that R--charges have to be constant on the conformal manifold. Of course in this case one has to know a priori that a fixed point with these R--charges exists.} After this linear analysis a--maximization was performed over the two dimensional space, leading to the exact values of the R--charges.

As explained in \cite{Herzog:2004tr, Benvenuti:2004wx}, the subset of the $3 p + q$--dimensional space of couplings preserving conformal invariance is two dimensional. This conformal surface is very similar to the one discussed in section \ref{section_margin} for the conifold \cite{kw}.

There is one marginal direction corresponding in supergravity to the axion--dilaton and one related to the vev of the complex B field on the $S^2$ (recall that for all $Y^{p, q}$ the topology is $S^2 \times S^3$). A difference is that the latter deformation is a little bit more complicated in terms of field theory couplings. For the conifold it simply corresponds to increasing one (inverse squared) gauge coupling and decreasing the other one by the same amount. For a general $Y^{p, q}$, one has also to change all the superpotential couplings in a non trivial way. As a consequence, there is no fixed line where all superpotential couplings vanish, as in Figure \ref{flowcon1}.

One can visualize the flow to the IR conformal manifold, starting from the free theory with matter as in Table \ref{examplefig}, in the following way. First the gauge couplings of the nodes with $N_F / N_C = 2$ (the ones adjacent to the external singlet $Z$--fields) may flow to the IR; in this way some bifundamentals have R--charge $1/2$. At this point some cubic superpotential couplings are relevant and can be sent to an interacting IR fixed point. Now it is possible to flow the gauge couplings of the nodes with $N_F / N_C = 3$ next to the previously ``activated'' nodes. Going on like this one reaches the conformal manifold, where all the couplings are non vanishing. Note that it is crucial that all the fields enter in the superpotential.

\subsection{R--charge 2 chiral operators and the $\bg$--deformation}

In the previous subsection the conformal surface over which the global symmetry is preserved has been described. In \cite{Benvenuti:2004wx} it was suggested that by breaking the $SU(2)$ global symmetry there is a possibility of finding additional marginal deformations. The simplest way of analysing this problem is to use the quotient procedure of \cite{Kol:2002zt}, as was done in section \ref{section_margin} for the case of the conifold. In order to do this, we have to find the set of chiral scalar operators with R--charge $2$. These operators are always single trace (the only exception could be the conifold) and are associated to close ``short'' loops of length $3$ or $4$ in the quivers, exactly as the operators entering the superpotential (\ref{WYpq}). They are generically of the form $\tr ( Z U Y U )$ or $\tr ( U V Y )$. All these operators have R--charge $2$, however, only a subset of them are chiral and have dimension $3$ on the whole IR fixed manifold. There are $p + q$ such loops: $2 q$ of length $3$ and $p - q$ of length $4$. Moreover, since the fields $U$ and $V$ transform in the spin--$1/2$ of the global $SU(2)$, all the loops transform in the spin $1/2 \otimes 1/2 = 0 \oplus 1$. We are thus dealing with a total of $4 ( p + q )$ operators. As we are going to show, the $\mF$-term relations imply that only $3$ of them are chiral. The equations of motion, derived from the superpotential (\ref{WYpq}), of the $Y$ and $Z$ fields
say that the spin--$0$ components vanish in the chiral ring. The equations of motion for the $U$ and $V$ doublet fields enable to `move' the loops around the quiver. All these loops are thus equal in the chiral ring. 

In conclusion there are precisely $3$ operators with R--charge $2$ and scaling dimension $\Delta = 3$; they transform in the spin--$1$ representation of $SU(2)$. Of course, the $U(1)_F$ charge of these operators is $0$, since $U(1)_F$ commutes with $SU(2)$.

The space of exactly marginal superpotential deformations is thus simply the $3$--complex dimensional spin--$1$ representation modded out by $SU(2, \IC)$. There is thus precisely one exactly marginal deformation, which can be given explicitly by the component with vanishing spin--$z$ (thus breaking $SU(2)$ to $U(1)_Z$): 
\beq\label{defYpq}
W_\bg \propto \sum_{i} \sigma_{\!3\, \ag}^{\;\;\bg} \; 
( U_{\!i}^\ag \, V_{\!i\,\bg} \, Y_{\!2i+2} + V_{\!i}^\ag \, U_{\!i+1\;\bg} \, Y_{\!2i+3} ) + 
\sum_{j} \sigma_{\!3\, \ag}^{\;\;\bg} \; Z_{\!j} \, U_{\!j+1}^\ag \, Y_{\!2j+3} \, U_{\!j\;\bg} \;.
\eeq
As in section \ref{section_margin} it is easy to verify the exact marginality using Leigh--Strassler arguments. The point is that the $\IZ_2$ symmetry
\beq
U_i^1 \leftrightarrow U_i^2 \;\;\;\;\; and \;\;\;\;\; V_i^1 \leftrightarrow V_i^2
\eeq
is still a symmetry of the full Lagrangian. This, together with the residual $U(1)^2$ continuos flavor symmetry, implies that the $2 \times 2$ matrices of scaling dimensions of the doublets are proportional to the identity.

We see that (\ref{defYpq}) is simply given by the toric superpotential, taken with all ``plus'' signs, and thus can be called $\bg$--deformation, as in the well known case of $\mN = 4$ SYM. This deformation is present for any superconformal field theory arising as the low energy field theory of D3 branes placed at the tip of a toric Calabi-Yau cone.

The general feature of the $\bg$--deformation is that the global symmetries are broken to their Cartan generators. This is a consequence of the fact that the $\bg$--deformed superpotential contains exactly the same operators as the undeformed toric superpotential. In the case of the $Y^{p, q}$, the flavor symmetry $SU(2) \times U(1)_F$, is broken to $U(1)_Z \times U(1)_F$. For generic superconformal quivers associated to toric Calabi--Yau cones, the continuos symmetry corresponding to isometries of the Sasaki--Einstein base is $U(1)^2_F \times U(1)_R$, and is preserved by the $\bg$--deformation.

In the case of $Y^{p, p}$ it is possible to check this computation against the results of \cite{Aharony:2002tp}, where the exactly marginal deformations were found, using methods different from ours, for the supersymmetric orbifolds $\IC^3 / \IZ_k$. Considering the case of $\IC^3 / \IZ_{2 p}$ with action $(1, 1, 2 p - 2)$, corresponding to $Y^{p, p}$, and keeping track of the constancy of the scaling dimensions on the conformal manifold, it seems that there is agreement with our result of a three--complex dimensional conformal manifold.

\pagebreak[3]

\subsection{Special cases and RG flows: $Y^{1, 1}\rightarrow Y^{1, 0}$ and $Y^{2, 2}\rightarrow Y^{2, 0}$}
The discussion in the above two paragraphs needs to be refined in some special cases.

For $Y^{1, 1}$ the Calabi--Yau cone is $\IC \times \IC^2 / \IZ_2$, so there is an accidental $\mN = 2$ supersymmetry, and one could expect additional marginal directions. The space of marginal deformations is $12$ dimensional: $2$ gauge couplings, $8$ superpotential couplings of the form $\tr ( Y U V )$ ($Y$ is now one of the two adjoints, $U$ and $V$ are still bifundamentals) and $2$ superpotential couplings of the form $\tr ( Y^3 )$. The naive global symmetry is the $10$ dimensional $U(2) \times U(2) \times U(1)^2$. The residual global symmetry is only the baryonic $U(1)$, so the quotient procedure tells us that the conformal manifold has dimension $12 - (10 - 1) = 3$, as in the generic $Y^{p, q}$ case. (Here we used the fact that for orbifolds the conformal manifolds passes over the free fixed point.)

$Y^{1, 0}$ is the conifold, and in section \ref{section_margin} a $5$ dimensional conformal manifold has been found. $Y^{2, 0} = Y^{1, 0} / \IZ_2$ is an orbifold of the conifold, but it is in a sense special, since $Y^{2, 0}$ is a regular Sasaki--Einstein $U(1)$--fibration over $\IC\IP^1 \times \IC\IP^1$, like $Y^{1, 0}$. The difference is that the length of the fiber is one half. The global flavor symmetry is thus $SU(2) \times SU(2)$: the fields $Y$ and $Z$ transform as a doublet of the ``accidental'' $SU(2)$. On a generic point of the symmetry preserving conformal surface, the space of R--charge $2$ chiral operators, instead of being $3$ dimensional, is $9$ dimensional, transforming in the spin--$(1, 1)$. Quotienting one finds $3$ marginal deformations associated to fluxes, corresponding to the deformations of the conifold. It is interesting to note that there is a line on the conformal manifold (where the superpotential vanishes, analog of Figure \ref{flowcon1}) where the global flavor symmetry is enhanced to $SU(2)^{4}$. Clearly these particular points cannot be described by supergravity.

The special cases $Y^{1, 1}$, $Y^{1, 0}$ and $Y^{2, 2}$ are also interesting as the only $Y^{p, q}$ admitting massive supersymmetric deformations. As is well known, adding mass terms to $Y^{1, 1}$ leads to $Y^{1, 0}$. Starting from $Y^{1, 1}$ without fluxes, it is possible to reach the conifold without fluxes through the Klebanov--Witten flow (twisted mass terms) and the Pilch--Warner deformation of the conifold through the Pilch--Warner flow (untwisted mass terms), which is a $\IZ_2$ orbifold of the flow $\mN=4\rightarrow\mN=1^*$. The $\bg$-deformation discussed in the previous paragraph generalizes to the whole flows; of course, the dual background is not known. 

As $Y^{1, 1}$, $Y^{2, 2}$ admits massive deformations of the form $\tr (Y Y)$, leading to $Y^{2, 0}$. Starting from $Y^{2, 2}$ without the $\bg$-deformation, one gets in the IR a superpotential analogous to the Pilch--Warner deformation of $Y^{1, 0}$. This flow is thus a $\IZ_2 \times \IZ_2$ orbifold of the flow $\mN=4 \rightarrow \mN=1^*$. In \cite{warner04} a large class of orbifolds, different from this one, of the flow $\mN=4 \rightarrow \mN=1^*$ has been studied.

\pagebreak[3]

\section{Moduli spaces of vacua and Higgsed theories for the conifold}
\label{section_moduli} 
In section \ref{section_margin} it was found that
the space of exactly marginal superpotentials is given by the
quotient $\mM_C = [(0, 0)_{\IC} \oplus (1, 1)_{\IC}]/ SO (4 ,
\IC)$; this has been verified with standard Leigh--Strassler arguments. In
this section
the moduli space of vacua of the gauge theory is discussed, with a
generic marginal superpotential of the form (\ref{genericW2}). It turns out that, in
order to study the moduli space $\mM_V$, it is possible to use the
complexified global symmetry group as well. The low energy Higgsed
theories for the various possibilities are also found.

\subsection{Generic quartic superpotentials}
Turning on superpotential deformations corresponds to giving vevs to the bosonic fields of Type IIB string theory. These fields effectively give a potential for a D3 brane probing the background. This potential is a function of the transverse coordinates and should be flat exactly on the gauge theoretical moduli space. Since we are interested in keeping the analogy with this simple brane probe we consider only vevs such that the four $N \times N$ matrices $A_1$, $A_2$, $B_{\dot{1}}$, $B_{\dot{2}}$ are diagonal. This corresponds to seeing only the symmetrized product part of the full moduli space of vacua $\mM_V$. Since the moduli space of the gauge theory should correspond to the configurations space of $N$ parallel D3 branes, we expect that the full $\mM_V$ is the symmetrized product of a smaller ``constituent space'' $Y$: \beq \mathcal{M}_V = Y^N/S_N\;\;, \eeq where $S_N$ is the symmetric group of permutations.\footnote{This is true in the case of the Klebanov-Witten superpotential \cite{Berenstein:2001uv}. For a generic point on the conformal manifold, there is the interesting possibility of finding some additional Higgsed vacua (with respect to the symmetrized product) given by non-diagonal bifundamentals. Such ``non-commutative'' vacua have been, for instance, analysed in \cite{Berenstein:2000ux, Dorey:2003pp, Dorey:2004iq, Hollowood:2004ek, Benini:2004nn, Dorey:2004xm} for various deformations of $\mN = 4$ SYM. In \cite{Dasgupta:2000hn} these issues have been studied for field theories living at conifold singularities.} 
For ease of discussion we can also take the vevs to be proportional to the identity matrix, seeing only the ``constituent space'' $Y$; this is what will be called moduli space of vacua \mc in the following. This \mc corresponds to the space seen by the stack of the $N$ coincident D3 branes taken away from the singularity to ``explore'' the transverse space.

In order to compute $\mM_V$, instead of modding out by the complexified gauge group, it is easier to work with gauge invariant variables and impose just $\mF$-term constraints. With the four diagonal bifundamental fields $A_1 , A_2 , B_{\dot{1}} , B_{\dot{2}}$ it is possible to construct $4$ gauge invariant ``mesons'', quadratic operators of the form $A_i\,B_{j}$: 
\beq x = A_1\,B_{\dot{1}}\,,\;\; y = A_2\,B_{\dot{2}}\,,\;\;
z = A_1\,B_{\dot{2}}\,,\;\; w = A_2\,B_{\dot{1}}\;. \eeq 
These operators parametrize the moduli space of vacua of the theory without superpotential. Of course, since the fundamental chiral fields are diagonal and commute with each other, the four mesons have to satisfy the relation $x\,y = z\,w$. This relation represents a complex cone over a quadric in $\IC\IP^3$, and indeed it coincides with (\ref{conifoldwi}), after a linear change of variables in the space $(x, y, z, w)$: 
\beq\label{changevariables}
\left\{
\begin{array}{l}
w_1 = \frac{1}{2} ( x + y ) \\
w_2 = \frac{i}{2} ( y - x ) \\
w_3 = \frac{i}{2} ( z + w ) \\
w_4 = \frac{1}{2} ( w - z )\;.
\end{array} \right. \eeq 
The result of this discussion is that the moduli space of the gauge theory we are considering, without superpotential, is the conifold. This is true everywhere on the fixed line of Figure \ref{flowcon1}, if both the gauge couplings are non-vanishing.

Adding a superpotential proportional to the Klebanov-Witten term (\ref{Wkw}) does not change this discussion, since for diagonal bifundamentals this term vanishes. So the moduli space is always the full $3$ dimensional conifold for a $SU(2) \times SU(2)$ preserving deformation. This is true on the $2$-complex dimensional manifold parametrized by the difference between the gauge couplings and by the superpotential Klebanov-Witten coupling.

Adding now a generic superpotential term in the $(1, 1)$ of $SU(2)
\times SU(2)$ generates more relations between the $4$ complex
variables $(x, y, z, w)$, coming from the usual $\mF$-flatness
relations \beq\label{genconstraint} \frac{\partial}{\partial
A_i}\mW = \frac{\partial}{\partial B_j}\mW = 0\;. \eeq We
immediately see that these moduli spaces of vacua are always
complex submanifolds of the $3$-dimensional conifold. Their
dimension can be $2$, $1$, or $0$. We will see that the last case
is the generic one: for generic $\lambda^{(\ag \bg) , \; (\agp
\bgp)}$ all flat directions are removed, the only supersymmetric
point allowed is the origin: $x = y = w = z = 0$. This fact is
expected, since the space of $\lambda^{(\ag \bg) , \; (\agp
\bgp)}$ is $9$ dimensional, so there should generically be more
than two independent relations from (\ref{genconstraint}). On the
other hand, it is obvious that the point $x = y = w = z = 0$ always
satisfies (\ref{genconstraint}), since the terms in
(\ref{genconstraint}) are cubic polynomials in the fundamental
fields.

In order to work with gauge invariant quantities we consider,
instead of (\ref{genconstraint}), the following gauge invariant
equations: \beq\label{ginvconstraint} A_{\ag} \;
\frac{\partial}{\partial A_{\bg}}\mW = B_{\dot{\ag}} \;
\frac{\partial}{\partial B_{\dot{\bg}}}\mW = 0 \,,\;\;\;\;\;\;
\textrm{for any} \;\; \ag, \bg, \dot{\ag}, \dot{\bg}\,. \eeq 
These are in total $8$ equations, not all independent.
Since the superpotential is a homogeneous quadratic function both
in the variables $A_{\ag}$ and in the variables $B_{\dot{\ag}}$,
the following relations hold for any $\mW$:
\beq\label{homogeneity} \left( A_{1} \frac{\partial}{\partial
A_{1}} + A_{2} \frac{\partial}{\partial A_{2}} \right) \mW =
\left(  B_{\dot{1}} \frac{\partial}{\partial B_{\dot{1}}} +
B_{\dot{2}} \frac{\partial}{\partial B_{\dot{2}}} \right) \mW = 2
\, \mW \eeq We are thus left with $7$ relations: the
supersymmetric condition $\mW = 0$ and other $6$ relations, from
(\ref{ginvconstraint}), independent from the $2$ listed in
(\ref{homogeneity}). All of these $7$ equations are quadratic in
the mesonic variables $(x, y, z, w)$.

 The $6$ differential
operators in (\ref{ginvconstraint}) independent from $\mW = 0$
transform in the $(1, 0) \oplus (0, 1)$ of $SU(2) \times SU(2)$.
The corresponding $6$ differential equations can thus be
rewritten, in the gauge invariant variables of
(\ref{changevariables}) $w_i$, as: \beq \mD_{i, j} \mW \equiv
\left( w_i \frac{\partial}{\partial w_j} - w_j
\frac{\partial}{\partial w_i} \right) \mW = 0 \;\;\;\;\;
\textrm{for all}\;i ,\, j \, = 1 \ldots 4\;. \eeq This is clear
from the symmetry of the problem, and can be verified using the
explicit change of variables (\ref{changevariables}). At this
point we can forget the original variables, and study the system
\beq\label{relationsinwi} \left\{
\begin{array}{l}
w_1^2 + w_2^2 + w_3^2 + w_4^2 = 0\\
\, \mW \, = \, 0\\
\mD_{i, j} \mW = 0\;.
\end{array}
\right. \eeq
We remind that the first equation in (\ref{relationsinwi}) is trivial, in the
sense that it follows directly from the definition of the $w_i$ (\ref{changevariables}),
and that $\mW$ in these variables is given by \beq \mW = \sum_{i,
j} M^{i, j} w_i w_j\,, \eeq where $M^{i, j}$ is a generic
traceless symmetric $4 \times 4$ complex matrix.

At this point it is clear that we can use a symmetry which is larger than
the previous $6$-real dimensional $SU(2) \times SU(2) = SO( 4 ,
\IR )$. It is possible to make a linear change of variables such that
the operators $\mD_{i, j}$ are invariant in form and the structure
$w_1^2 + w_2^2 + w_3^2 + w_4^2$ does not change. This change of
variables constitute the full $SO (4, \IC)$ group of rotations in
$\IC^4$. Using this redefinition freedom we can thus put the
matrix $M$ in a canonical, possibly simple, form. Since the matrix
is complex it is not always possible to diagonalize it. We
consider in this section the cases in which $M$ is diagonalizable,
corresponding to stable orbits of the quotient \mc $= (1, 1)_{\IC}
/ SO (4 , \IC)$. In the next section we then complete the
discussion for the ``unstable'' superpotentials, when $M$ is not
diagonalizable.

With the traceless matrix $M^{i, j}$ in a diagonal form with
eigenvalues $\lambda_k$, performing the derivatives in $\mD_{i, j}
\mW$, the problem is reduced to 
\begin{subnumcases}{}
 \sum_k \, \, w_k^2 = 0 \label{const1}\\
 \sum_k \, \lambda_k \, w_k^2 = 0 \label{const2}\\
 \left( \lambda_i - \lambda_j \right) w_i w_j = 0\;, \label{const3}
\end{subnumcases}
for all the possible choices of the four
complex numbers $\lambda_k$ satisfying $\sum_k \lambda_k = 0$.

In the generic case all $\lambda_i$ are non-equal: (\ref{const3}) implies that at
least three out of the four $w_i$ vanish. With this assumption
however (\ref{const1}) says that all of the $4$ $w_i$ are zero.
This is the expected result that for a generic superpotential
there are no flat directions.

Non generic diagonalizable superpotentials correspond to
situations in which some of the $\lambda_i$ are equal. There are
four different possibilities, and we go on to discuss them case by
case.

\begin{itemize}
\item[\bf{A}]{\large $\phantom{aaaa}\lambda_1 = \lambda_2 = \lambda_3 = \lambda_4$\,.}\\
(The relation $\sum_k \lambda_k = 0$ implies that all $\lambda_k$s
vanish). This is the Klebanov-Witten superpotential and the moduli
space is the full conifold. The global flavor symmetry is $SU(2)
\times SU(2)$ and acts on the base $\IP^1 \times \IP^1$. As in
\cite{kw, Klebanov:1999tb} one can consider the low energy
Higgsed theory. Giving vevs proportional to the identity matrix to
the bifundamental fields corresponds to pulling the stack of D3
branes out of the singularity, so the low energy
theory has to be $\mN = 4$ SYM with group $SU(N)$. This can be seen
explicitly by giving a vev to the fundamental fields: at low
energy there is a cubic superpotential which reproduces the $\mN =
4$ superpotential and a quartic term which is irrelevant.

\item[\bf{B}]{\large $\phantom{aaaa}\lambda_1 = \lambda_2 = \lambda_3 \neq \lambda_4$\,.}\\
This case corresponds to the Pilch-Warner deformation.
(\ref{const3}) simplifies to \beq w_4 w_1 = w_4 w_2 = w_4 w_3 =
0\,. \eeq $w_4 \neq 0$ implies $w_1 = w_2 = w_3 = 0$, but with
these restrictions both (\ref{const1}) and (\ref{const2}) become
$w_4^2 = 0$. So $w_4 = 0$: (\ref{const3}) are automatically
satisfied, while both (\ref{const1}) and (\ref{const2}) become
 \beq\label{msPW}
  w_1^2 + w_2^2 + w_3^2 = 0\,.
 \eeq
 The moduli space is a singular $2$ dimensional complex manifold $\IC^2/\IZ_2$: a complex
cone over $\IP^1$. The deformation explicitly is given by
 \beq \mW\label{WexplPW}
= \mW_{KW} + \lambda_{PW} \, tr \left( A_{1} B_{\dot{1}} A_{2}
B_{\dot{2}} + A_{1} B_{\dot{2}} A_{2} B_{\dot{1}} -  A_{1}
B_{\dot{2}} A_{1} B_{\dot{2}} - A_{2} B_{\dot{1}} A_{2}
B_{\dot{1}} \right)
 \eeq
 The global flavor symmetry is broken to
$SU(2)$, coming from a diagonal combination of the previous $SU(2)
\times SU(2)$. As in the previous case, all possible vevs are
equivalent (modulo an overall complex rescaling) thanks to the
global symmetry $SU(2)$, which acts on the base $\IP^1$. The low
energy Higgsed theory has to be the theory of D3 branes at a
smooth point with some fluxes turned on. Moreover, the dimension
of \mv of this deformed $\mN = 4$ SYM has to be exactly two. There
is only one deformation of this type, corresponding to the
addition of one supersymmetric mass term. Let us reproduce
explicitly this result. The following vevs to the bifundamentals
 \beq\label{vevsPW} \left\{
\begin{array}{l}
\langle \, A_1 \, \rangle = \langle \, B_{\dot{1}} \, \rangle = v \, \mathbf{1}_{N \times N} \\
\langle \, A_2 \, \rangle = \langle \, B_{\dot{2}} \, \rangle = 0 
\end{array}
\right.
 \eeq
satisfy both the $\mF$-term constraints (\ref{msPW}) and the
$\mD$-term relation 
\beq
|A_1|^2 + |A_2|^2 = |B_{\dot{1}}|^2 + |B_{\dot{2}}|^2 \;.
\eeq
Substituting $A_i = \langle A_i\rangle + \ag_i$ and $B_i = \langle B_i \rangle + \bg_i$
in (\ref{WexplPW}) one obtains
 \bea \label{PW+vev}
 \mathcal{W}_{Higgsed} = & v \lambda_{KW} \tr \left( (\ag_1 +
 \bg_1) [ \ag_2 , \, \bg_{2} ] \right)
  - 2 v^2 \lambda_{PW} \, tr \left( \ag_{2} + \bg_2 \right)^2 \\
 & + v \lambda_{PW} \, \tr \left( (\ag_1 + \bg_1) \{ \ag_2 , \, \bg_2 \}
  - 2 \ag_1 \bg_2^2 - 2  \bg_1 \ag_2^2 \right) \nonumber \;.
 \eea
The term proportional to $v^2 \lambda_{PW}$ is a mass term and
becomes exactly marginal in the IR. As a consequence the term in the
last line of \ref{PW+vev}, proportional to $v \lambda_{PW}$ at the
IR fixed point is irrelevant and has to be neglected. The low
energy Higgsed theory is thus $\mN = 4$ SYM deformed by one mass
term, the $\mN = 1^*$. This theory has been studied holographically in \cite{Freedman:1999gp, Pilch:2000ej, Pilch:2000fu, Pilch:2004yg}. In \cite{orbifoldsugra} it has been argued that the conifold with the Pilch-Warner superpotential is a $\IZ_2$ orbifold of the $\mN = 1^*$ theory, consistent with our results.

\item[\bf{C}]{\large $\phantom{aaaa}\lambda_1 = \lambda_2 \neq \lambda_3 = \lambda_4$\,.}\\
This is the analog of the so called $\bg$-deformation of $\mN =4$
SYM. In this case (\ref{const3}) reads
 \beq w_1 w_3 = w_1 w_4 =
w_2 w_3 = w_2 w_4 = 0\,. \eeq
 The previous equations are solved by $w_1 = w_2 = 0$ or $w_3 = w_4 = 0$. In this case both
(\ref{const1}) and (\ref{const2}) become \beq w_1 = w_2 = 0 \,,\;
w_3^2 + w_4^2 = 0 \;\;\;\;\; \textrm{or} \;\;\;\;\; w_3 = w_4 =
0\,,\; w_1^2 + w_2^2 = 0\,. \eeq The moduli space thus consists of
four distinct one-dimensional branches intersecting at the same
point. The explicit form of the superpotential is
 \beq \mW = \mW_{KW} + \lambda_{\bg} \,
\tr \left( A_{1} B_{\dot{1}} A_{2} B_{\dot{2}} +  A_{1} B_{\dot{2}}
A_{2} B_{\dot{1}} \right)\,.
 \eeq


In this case the residual flavor symmetry corresponds to the
Cartan generators of the original $SU(2) \times SU(2)$, acting as
 \beq
 A_1 \rightarrow e^{i \ag} A_1 \;, \;\;A_2 \rightarrow e^{- i \ag} A_2
\;, \;\; B_{\dot{1}} \rightarrow e^{i \bg} B_{\dot{1}} \;, \;\;B_{\dot{2}} \rightarrow
e^{- i \bg} B_{\dot{2}} \;.
 \eeq
This is a general feature of the ``$\bg$-deformations" of toric
superpotentials: the non--Abelian flavor symmetries are broken to
their Cartan generators. Note that one $U(1)$ factor acts on two
branches of $\mM_V$, the other $U(1)$ factor acts on the other two
branches.

Giving vevs as in the previous paragraph, one finds that the low
energy Higgsed theory is pure $SU(N)$ $\mN = 2$ SYM, as in $\mN =
4$ with $2$ mass terms added. We note that this does not imply that the present deformation corresponds to an orbifold of the supergravity solution associated to pure $\mN = 2$ SYM.

\item[\bf{D}]{\large $\phantom{aaaa}\lambda_1 = \lambda_2$, $\lambda_3$ and $\lambda_4$ generic.}\\
This deformation contains as special limits the Pilch-Warner deformation {\bf B} and the $\bg$--deformation {\bf C}. Here all products $w_i w_j$ vanish except $w_1 w_2$. As in case {\bf B},
$w_4$ has to vanish, and for the same reason $w_3$ as well. As a
consequence both (\ref{const1}) and (\ref{const2}) become $w_1^2 +
w_2^2 = 0$, and the moduli space consists of two distinct
one-dimensional branches intersecting at the origin: 
\beq
 w_3 = w_4 = 0\,,\; w_1^2 + w_2^2 = 0\;\; \longleftrightarrow  \;\; x y = z = w = 0\;.
\eeq
 The explicit superpotential is
 \beq
  \mW = \mW_{KW} + \lambda_1 \, \tr \left(
A_{1} B_{\dot{1}} A_{2} B_{\dot{2}} +  A_{1} B_{\dot{2}} A_{2}
B_{\dot{1}} \right) + \lambda_2 \, tr \left( A_{1} B_{\dot{2}}
A_{1} B_{\dot{2}} + A_{2} B_{\dot{1}} A_{2} B_{\dot{1}} \right)
 \eeq

The residual symmetry here is $U(1)$: $A_1 \rightarrow e^{i \ag}
A_1 \;, \;\;A_2 \rightarrow e^{- i \ag} A_2 \;, \;\; B_{\dot{1}}
\rightarrow e^{i \ag} B_{\dot{1}} \;, \;\;B_{\dot{2}} \rightarrow e^{- i \ag} B_{\dot{2}}
\;,$ giving vevs to $A_1$ and $B_{\dot{1}}$ (or $A_2$ and $B_{\dot{2}}$) one still
finds pure $\mN = 2$ SYM.

\end{itemize}

\subsection{Non diagonalizable superpotentials are marginally irrelevant}
We can now proceed to consider what happens when the matrix $M$ is
not diagonalizable. These matrices are unstable orbits of the
complexified global symmetry group, since for a non diagonalizable
matrix at least two eigenvectors coincide, so a small deformation
resolves this degeneracy and makes the matrix diagonalizable. Also
in this case it is possible to give an explicit classification of
the different possible superpotentials. We now discuss a simple
example, then we analize all the classification. The results are
always similar.

Let us consider the following superpotential:
\beq \label{a2b2a2b2}
\mathcal{W} = \lambda_{KW} \, \tr \left( A_{1} B_{\dot{1}} A_{2} B_{\dot{2}} - A_{1} B_{\dot{2}} A_{2} B_{\dot{1}} \right) + \eg \, \tr \left( A_{2} B_{\dot{2}} A_{2} B_{\dot{2}} \right) \;.
\eeq
This corresponds to a matrix $M_{i j}$ of the following form:
\begin{small}\beq\label{2x2vanish}
M =  \eg \, \left(
\begin{array}{cccc}
    i       &     1     &   0   &    0       \\
    1       &    - i    &   0   &    0       \\
    0       &    0      &   0   &    0       \\
    0       &    0      &   0   &    0
\end{array}
\right) \, . \eeq\end{small} Since both the trace and the
determinant of the non-vanishing $2 \times 2$ submatrix are zero,
this matrix is non diagonalizable.

The deformation $\tr \left( A_{2} B_{\dot{2}} A_{2} B_{\dot{2}}
\right)$ breaks all the flavor symmetries, so the R--charge
$r_A$ and $r_B$ are in general expected to be $2 \times 2$
matrices. Since we have just four couplings and, in order to have
an RG fixed point, we have to impose more than $4$ constraints,
Leigh--Strassler arguments do not work, and it seems unlikely that
the deformation we are considering is exactly marginal. We now
provide an argument that implies that $\tr \left( A_{2} B_{\dot{2}}
A_{2} B_{\dot{2}} \right)$ is a marginally \emph{irrelevant}
deformation of the strongly interacting conifold field theory.

Let us suppose $\tr \left( A_{2} B_{\dot{2}} A_{2} B_{\dot{2}}
\right)$ is marginally relevant or exactly marginal. It is in this
case possible to compute, as in the previous section, the moduli
space of vacua \mv. This moduli space \mv is at most $2$ complex
dimensional, since there is at least one independent $\mF$-term
non trivial constraint from (\ref{a2b2a2b2}).\footnote{The results
of the following analysis is that it would actually be $2$-complex
dimensional, but we don't need this for our argument.} Since
all the $\mF$-term equations are proportional to $A_2 B_{\dot{2}}$, it is
clear that the same vevs as in (\ref{vevsPW}) satisfy both $\mD$ and $\mF$-terms
contraints:
 \beq \left\{ \begin{array}{l}
\langle \, A_1 \, \rangle = \langle \, B_{\dot{1}} \, \rangle = v \, \mathbf{1}_{N \times N} \\
\langle \, A_2 \, \rangle = \langle \, B_{\dot{2}} \, \rangle = 0 
\end{array}\right. \eeq
What is the low energy field theory with these vevs?
Substituting as in the previous section $A_i = \langle A_i \rangle + \ag_i$
and $B_i = \langle B_i \rangle + \bg_i$ in (\ref{a2b2a2b2}) one obtains, 
\beq
\label{a2b2a2b2+vev} \mathcal{W}_{Higgsed} = v \lambda_{KW} \tr
\left( (\ag_1 + \bg_1) [ \ag_2 , \, \bg_2 ]
\right) + \eg \, \tr \left( \ag_{2} \bg_2 \ag_{2}
\bg_2 \right) \;. \eeq 
This is an irrelevant deformation
of the $\mN = 4$ theory, so at low energies we find the pure $\mN
= 4$ theory, which has a $3$ dimensional space of vacua. The
theory described by (\ref{a2b2a2b2}) has a \mv with a smaller
dimension (assuming that the deformation proportional to $\eg$ is
not irrelevant). This is a contradiction, since the IR \mv should
have a dimension equal to, or smaller than, the UV \mv. The
conclusion is that $\tr \left( A_{2} B_{\dot{2}} A_{2} B_{\dot{2}}
\right)$ is a marginally \emph{irrelevant} deformation; at low
energy only the term proportional to $\lambda_{KW}$ survives.

This results completes the analogy between \mv and \mc. In order
to compute \mv one has to neglect unstable orbits, and here we
saw that a non diagonalizable matrix $M$ also has to be
neglected, since it corresponds to an irrelevant deformation.

The result obtained for the matrix (\ref{2x2vanish}) can be
immediately generalized to the case of a non diagonalizable matrix
of the following form:
\begin{small}\beq\label{2x2gen}
M = \eg \, \left(
\begin{array}{cccc}
 \lambda + i& 1         &   0                 &    0       \\
    1       &\lambda - i&   0                 &    0       \\
    0       &    0      &- \lambda + \lambda_1&    0       \\
    0       &    0      &   0                 &- \lambda - \lambda_1
\end{array}
\right)\phantom{aaaaaa}\lambda , \; \lambda_1 \;\;
\textrm{generic.} \eeq\end{small} The reason is clear: if a
deformations corresponding to (\ref{2x2gen}) is marginal or
relevant (possibly for a particular combinations of $\lambda$ and
$\lambda_1$), then also in the limit $\lambda , \,\lambda_1
\rightarrow 0$ this deformation is marginal or relevant. But we
just showed that in this limit the deformation is irrelevant.
Thus, all deformations corresponding to (\ref{2x2gen}) are
irrelevant.

It is now possible to go on and discuss the case of general non diagonalizable
matrices. Since we were not able to find a general argument, a boring case by case analysis will be performed. The result is that
they always lead to marginally irrelevant terms. We relegate this discussion in the Appendix \ref{nondiago}.

\section{Comments}
\label{section_conclu}
We have shown that, for every specific model, the set of theories for which the AdS/CFT duality is at work is a non-trivial manifold of superconformal field theories.

One marginal direction is generated by the ``standard'' lagrangian, and it corresponds on the gravity side to the vev of the axion-dilaton. Other marginal directions are obtained by giving a vev for the complex B-field over the $S^2$. The number of deformations of this type is precisely the second Betti number of the real $5$-dimensional base of the Calabi-Yau singularity. These two types of deformations are well known, and are peculiar in the sense that they leave the global symmetry and the moduli space of vacua untouched. There can also be purely geometric marginal deformations if there is a moduli space of Einstein metrics, as is the case for the del Pezzo$_n$ models, with $n > 4$ \cite{Wijnholt:2002qz}.

Analysing the set of quiver theories called $Y^{p, q}$, we have shown that there is always one more marginal direction. This additional marginal direction is generic for toric quivers and, in the case of $\mN = 4$ SYM, is known with the name of $\bg$-deformation.

In the case of the $T^{1, 1} = Y^{1, 0}$ and $T^{1, 1} / \IZ_2 = Y^{2, 0}$ there are two more marginal deformations. When only the Pilch-Warner superpotential is turned on, the dual Type IIB supergravity background is known explicitly. It is a $\IZ_2$ or $\IZ_2 \times \IZ_2$ orbifold of the $\mN = 1^*$ theory, whose complete dual 10D supergravity background has been constructed in \cite{Pilch:2000ej}, uplifting the results obtained in 5D gauged supergravity \cite{Freedman:1999gp}. To be more precise, the whole dual RG flow between the $\mN = 4$ SYM and the $\mN = 1^*$ theory is known \cite{Freedman:1999gp, Pilch:2000fu, Halmagyi:2004jy}. Orbifolding this flow one gets the dual RG flows $Y^{1, 1}\rightarrow Y^{1, 0}$ \cite{orbifoldsugra, corrado04, Halmagyi:2004jy} and $Y^{2, 2}\rightarrow Y^{2, 0}$. On the gauge theory side these supersymmetric flows are driven by untwisted sector mass terms.

\vspace{0.1 cm}

We think it would be very interesting to find the supergravity dual of the $\bg$-deformation for the $Y^{p, q}$ geometries. On the one hand it seems difficult to work in a $5$ dimensional supergravity perspective. In the homogeneous case of the conifold, the authors of \cite{Ceresole:1999zs} have been able to find the full Kaluza-Klein reduction to $5$ dimensions. For general $Y^{p, q}$ this has not been done. In any case, as the results of \cite{aharony2002} suggest, it is unlikely that a consistent truncation like $\mN = 8$ gauged supergravity can be used to discuss the marginal $\bg$-deformation. So it seems that the full 10D Type IIB supergravity has to be taken into consideration.

On the other hand, the $\bg$-deformation breaks only part of the global symmetry: for every $\bg$-deformed $Y^{p, q}$, and also for $S^5$ and all the other toric Sasaki--Einstein manifolds, the remaining isometry of the compact five manifold is $U(1)^3$. So the corresponding metric (which is not Einstein, due to the presence of fluxes) is expected to be cohomogeneity--two, meaning that it depends non trivially on two coordinates. This simplification occurs also for the fluxes. Since supersymmetry is clearly preserved, it is conceivable that the techniques of $G$--structures could lead to the determination of the explicit Type IIB supergravity compactification with fluxes.


\section*{Acknowledgements}
We wish to acknowledge Damiano Anselmi, Stefano Bolognesi, Chris Herzog, Barak Kol and Carlos Nu\~nez for useful discussions.
S.B. would like to thank the organizers of the 2004 Onassis Lectures on ``Fields and Strings'', where the lectures given by professor Igor Klebanov interested him in this problem.
We are grateful to Igor Klebanov also for comments on a draft version of this work.

Research supported in part by the CTP and the LNS
of MIT and the U.S. Department of Energy under cooperative agreement
$\#$DE-FC02-94ER40818, and by BSF American--Israeli Bi--National Science Foundation. A. H. is also supported by a DOE OJI award.


\appendix

\section{Generic non diagonalizable superpotentials}\label{nondiago}
\vspace{0.4 cm}
It is well known that every complex matrix, even if not
diagonalizable, can be put in the Jordan canonical form (of
course, if the original matrix is symmetric and non
diagonalizable, the change of basis is not a rotation). We now
show that it is actually possible to put any Jordan block in a
symmetric form with a $GL (n , \IC )$ change of basis. We
explicitly write down the matrices which change the basis. Given a Jordan
block $\mJ$
\begin{small}
\beq
\mJ = \left(
\begin{array}{ccccc}
    0     &  1   &  0   & \dots  &  \;   0   \\
    0     &  0   &  1   &        &   \;   0   \\
    0     &  0   &  0   &\ddots  &    \vdots  \\
 \vdots   &\vdots&      &\ddots  &    \;  1     \\
    0     &  0   &      &\dots   &   \;   0
\end{array}
\right)
\eeq
\end{small}
and the change of basis matrix $\mT$ defined by
\begin{small}
\beq
\mT = \left(
\begin{array}{cccc}
    1     &  0   & \dots  &     0     \\
    0     &  1   &        &     0   \\
  \vdots  &      &\ddots  &  \vdots \\
    0     &  0   &        &     1
\end{array}
\right)
- i
\left(
\begin{array}{cccc}
    0     & \dots  &   0    &     1     \\
    0     &        &   1    &     0   \\
 \vdots   & \cdot  &        &   \vdots \\
    1     &        &   0    &     0
\end{array}
\right)\,,
\eeq
\end{small}
\noindent it is straightforward to verify that $\mS = \mT \, \mJ
\, \mT^{- 1}$ is indeed symmetric ($\mT^{- 1}$ is related to the
complex conjugate of $\mT$ as $\mT^{- 1} = \mT^{*} / 2$).
Explicitly the result for what can be called the ``symmetrized''
Jordan block $\mS$ is
\begin{small}
\beq\label{symmJB}
2 \, \mS = \left(
\begin{array}{ccccc}
    0     &  1   &   0    &\dots  &    0     \\
    1     &  0   &   1    &       & \cdot   \\
    0     &  1   & \cdot  &       &   0   \\
  \cdot   &\cdot & \cdot  &  0    &   1    \\
    0     &  0   &    0   &  1    &    0
\end{array}
\right)
+ i
\left(
\begin{array}{ccccc}
    0     & \dots  &   0    & 1   &   0     \\
    0     &        &   1    & 0   &  - 1   \\
    0     & \cdot  & \cdot  & - 1 &   0   \\
    1     &    0   & \cdot  &     & \cdot \\
    0     &   - 1  &   0    &  0  &   0
\end{array}
\right)\,.
\eeq
\end{small}
We note that in the case of $2 \times 2$ matrices (\ref{symmJB})
precisely coincides with the non vanishing part of
(\ref{2x2vanish}).

At this point we can use the following fact from matrix theory: if
two symmetric complex matrices are similar they are also
orthogonally similar, i.e. the similarity matrix can be chosen to
lie in $SO (n , \IC )$. The consequence is that any symmetric
complex matrix can be put in a standard ``symmetrized'' Jordan
block diagonal form, with a rotation. (This discussion also shows
that any complex matrix is similar to a symmetric matrix).

We can thus proceed and discuss the various possibilities that can
be obtained if the matrix $M$ is not diagonalizable. Since $M$ has
dimension $4$ there can be two blocks $2 \times 2$, one block $3
\times 3$ or one block $4 \times 4$. (The case of one block $2
\times 2$ has already been analysed).

\subsection*{$\phantom{aaaaaa}$ Two $2 \times 2$ ``symmetrized'' Jordan blocks}
From formula (\ref{symmJB}) we see that we have to analyse a
deformation of the following form:
\begin{small}\beq
M = \eg \, \left(
\begin{array}{cccc}
    i       &    1      &   0         &    0       \\
    1       &  - i      &   0         &    0       \\
    0       &    0      &   i         &    1       \\
    0       &    0      &   1         &    - i
\end{array}
\right)\phantom{aaaaaa} \eeq\end{small} Explicitly we are
considering a superpotential of the following form:
\beq\label{W2x2+2x2} \mW = \mW_{KW} + \eg \, tr \left( A_{2}
B_{\dot{2}} A_{2} B_{\dot{2}} -  A_{1} B_{\dot{2}} A_{1}
B_{\dot{2}} \right) \eeq Out of the $6$ equations $\mD_{i, j} =
0$, the two equations coming from $\mD_{1, 2}$ and $\mD_{3, 4}$
are a perfect square: \beq\left\{\begin{array}{l}
  \mD_{1, 2} \mW = ( w_1 - i w_2 )^2 \\
  \mD_{3, 4} \mW = ( w_3 - i w_4 )^2 \,.
\end{array}\right. \eeq
This implies $w_1 = i w_2$ and $w_3 = i w_4$. With this
requirement the other four equations of the form $\mD_{i, j} \mW =
0$ turn out to vanish trivially. Also the relations $\mW = 0$ and
$w_1^2 +w_2^2 +w_3^2 + w_4^2 = 0$ are clearly satisfied. We
conclude that the hypothetical moduli space is, topologically,
$\IC^2$: \beq w_1 = i w_2\,,\;\;\; w_3 = i w_4\,. \eeq A possible
vev satisfying $\mD$ and $\mF$-term constraints is as follows:
\beq\label{vevs2x2+2x2} \left\{\begin{array}{l}
\langle A_1 \rangle = \langle A_2 \rangle = v \\
\langle B_{\dot{1}} \rangle = \sqrt{2} \, v \\
\langle B_{\dot{2}} \rangle = 0 \; .
\end{array}\right.\eeq
The low energy theory after this Higgsing as usual has a $SU(N)$
gauge group, three adjoints and the following superpotential: \beq
\label{2x2+2x2+vev} \mathcal{W}_{Higgsed} = v \lambda_{KW} tr
\left( X [ Y , Z ] \right) + 2 \eg v \, tr \left( X \, Z^2 \right)
\;. \eeq Where $X = A_1 - A_2$, $Y = B_{\dot{1}} + \sqrt{2}$, $Z = B_{\dot{2}}$.
The term proportional to $\eg$ is a marginally irrelevant
deformation of $\mN = 4$ theory, which has a $3$ dimensional space
of vacua. So also in this case we find the same inconsistency as
before, and this deformation is irrelevant as well.

\subsection*{$\phantom{aaaaaa}$ One $3 \times 3$ ``symmetrized'' Jordan block}
Now the matrix can be put in the following form:
\begin{small}\beq
M \propto \, \left(
\begin{array}{cccc}
   0    & 1 + i &   0   &   0       \\
  1 + i &  0    & 1 - i &   0       \\
    0   & 1 - i &   0   &   0       \\
    0   &   0   &   0   &   0
\end{array}\right)\,,\eeq\end{small}
which corresponds to a superpotential \beq\label{W3x3} \mW =
\mW_{KW} + \eg \, tr \left( A_{1} B_{\dot{1}} A_{1} B_{\dot{1}} +
A_{1} B_{\dot{1}} A_{2} B_{\dot{1}} + A_{1} B_{\dot{1}} A_{1}
B_{\dot{2}} - A_{1} B_{\dot{2}} A_{2} B_{\dot{2}} -  A_{2}
B_{\dot{1}} A_{2} B_{\dot{2}} - A_{2} B_{\dot{2}} A_{2}
B_{\dot{2}} \right) \,. \eeq As before we exibit a couple of
equations coming from $\mD_{i, j} = 0$: \beq \left\{
\begin{array}{l}
(  \mD_{1, 2} + i \mD_{2, 3} ) \mW = ( 1 + i ) ( w_1 - i w_3 )^2 \\
  \mD_{2, 3} \mW = ( 1 - i ) (w_2^2 - i w_3 ( w_1 - i w_3 )) \,.
\end{array}
\right. \eeq The vanishing of the two previous expressions implies
$w_1 = i w_3$ and $w_2 = 0$. The final result, adding the
requirement $w_1^2 +w_2^2 +w_3^2 + w_4^2 = 0$, is that all the
constraints are satisfied by \beq w_1 = i w_3\,,\;\;\; w_2 = w_4 =
0\,, \eeq leading to a one dimensional \mv. A possible vev
satisfying $\mD$ and $\mF$-term constraints is  
$\langle A_1 \rangle = \langle B_{\dot{1}} \rangle = - \langle A_2 \rangle = - \langle B_{\dot{2}} \rangle = v$.
Defining $X = A_1 + A_2$, $Y = B_{\dot{1}} + B_{\dot{2}}$, $Z = A_2 + B_{\dot{2}}$, one
finds a Higgsed superpotential of the following form: \beq
\label{3x3+vev} \mathcal{W}_{Higgsed} = v \lambda_{KW} tr \left( X
[ Z , Y ] \right) + \eg v \, tr \left( X \, Y^2 + Y \, X^2 \right)
\;. \eeq The term proportional to $\eg$ is again a marginally
irrelevant deformation of $\mN = 4$ theory, so there is the same
inconsistency.

\subsection*{$\phantom{aaaaaa}$ One $4 \times 4$ ``symmetrized'' Jordan block}
This is the last case.
\begin{small}\beq
M \propto \left(
\begin{array}{cccc}
    0   &  1   &  i  &  0  \\
    1   &  i   &  1  & - i \\
    i   &  1   & - i &  1  \\
    0   & - i  &  1  &  0
\end{array}
\right)
\eeq\end{small}
In terms of the bifundamental fields, the deformation is proportional to
\beq
\begin{array}{r l}
\mW = \mW_{KW} + & \eg \, tr \big(
        3 A_2 B_{\dot{1}} A_2 B_{\dot{1}} + 2 i A_1 B_{\dot{1}} A_2 B_{\dot{1}} - 3 A_1 B_{\dot{1}} A_1 B_{\dot{1}} + 2 i A_2 B_{\dot{1}} A_2 B_{\dot{2}} + 2 A_1 B_{\dot{1}} A_2 B_{\dot{2}} +  \\
  &     A_2 B_{\dot{2}} A_2 B_{\dot{2}} + 2 A_1 B_{\dot{2}} A_2 B_{\dot{1}} - 2 i A_1 B_{\dot{1}} A_1 B_{\dot{2}} + 6 i A_1 B_{\dot{2}} A_2 B_{\dot{2}} - A_1 B_{\dot{2}} A_1 B_{\dot{2}} \big) \, .
\end{array}
\eeq 
Performing the derivatives in $\mD_{i, j} \mW = 0$ one finds that two linear
combinations of them can be written as \beq \left\{
\begin{array}{l}
( \mD_{1, 2} + i \mD_{1, 3} ) \mW = - ( w_2 + i w_3 )^2 + 2 w_1 ( w_3 + i w_2 )\\
( \mD_{1, 4} + \mD_{2, 3} )\mW = - ( w_3 + i w_2 )^2 \,.
\end{array}
\right. \eeq These two equations imply $w_2 = w_3 = 0$. It is
straightforward to check that all the other conditions in
(\ref{relationsinwi}) are verified by \beq\label{mv4x4} w_1 = i
w_4\,,\;\;\; w_2 = w_3 = 0\,. \eeq Giving the vevs as $\langle A_1 \rangle = \langle
B_{\dot{1}} \rangle = i \langle A_2 \rangle = - i \langle B_{\dot{2}} \rangle = v$, and defining $X = A_1 - i A_2$, $Y = B_{\dot{1}} +i B_{\dot{2}}$, $Z = A_2 + B_{\dot{2}}$, one finds the following Higgsed superpotential
\beq \label{4x4+vev} \mathcal{W}_{Higgsed} = v \lambda_{KW}
tr \left( X [ Z , Y ] \right) - 4 \eg v^2 \, tr \left( Y^2 \right)
- 4 \eg v \, tr \left( Y X^2 \right)\;. \eeq 
Modulo a marginally
irrelevant term, in the IR there is the $\mN = 1^*$ theory, which
has a $2$ dimensional moduli space. Also here there is an
inconsistency, since the \mv described by (\ref{mv4x4}) is $1$
dimensional.

\pagebreak[3]

\bibliographystyle{JHEP}

\end{document}